\documentclass[sigconf]{acmart}

\usepackage{amsmath}
\usepackage{booktabs,multirow}
\usepackage{graphicx}
\usepackage{microtype}
\usepackage{xcolor}
\usepackage{algorithm,algorithmic}
\usepackage{hyperref}

\setcopyright{acmcopyright}
\copyrightyear{2018}
\acmYear{2018}
\acmDOI{XXXXXXX.XXXXXXX}

\acmISBN{978-1-4503-XXXX-X/18/06}

\begin{document}

\title{Not Only NTP: Extending Training Signal Coverage for Generative Recommendation}

\author{Changhao Li}
\authornote{Corresponding author.}
\affiliation{%
  \institution{Meituan}
  \city{Chengdu}
  \country{China}}
\email{lichanghao@meituan.com}

\author{Shuli Wang}
\authornotemark[1]
\affiliation{%
  \institution{Meituan}
  \city{Chengdu}
  \country{China}}
\email{wangshuli03@meituan.com}

\author{Junwei Yin}
\affiliation{%
  \institution{Meituan}
  \city{Chengdu}
  \country{China}}
\email{yinjunwei03@meituan.com}

\author{Senjie Kou}
\affiliation{%
  \institution{Meituan}
  \city{Chengdu}
  \country{China}}
\email{kousenjie@meituan.com}

\author{Yinqiu Huang}
\affiliation{%
  \institution{Meituan}
  \city{Chengdu}
  \country{China}}
\email{huangyinqiu@meituan.com}

\author{Chi Wang}
\affiliation{%
  \institution{Meituan}
  \city{Chengdu}
  \country{China}}
\email{wangchi06@meituan.com}

\author{Yinhua Zhu}
\affiliation{%
  \institution{Meituan}
  \city{Chengdu}
  \country{China}}
\email{zhuyinhua@meituan.com}

\author{Haitao Wang}
\affiliation{%
  \institution{Meituan}
  \city{Chengdu}
  \country{China}}
\email{wanghaitao13@meituan.com}

\author{Xingxing Wang}
\affiliation{%
  \institution{Meituan}
  \city{Beijing}
  \country{China}}
\email{wangxingxing04@meituan.com}

\begin{abstract}
Next-Token Prediction (NTP) carries two structural training signal limitations. First, NTP optimizes for single-step prediction only, placing no supervised pressure on learning longer-range behavioral structure---we term this \textbf{temporal locality}. Second, in multi-domain sequences, each target item embedding receives gradient updates exclusively from the immediately preceding hidden state, with no explicit gradient pathway from cross-domain context---we term this \textbf{spatial locality}.

We propose \textbf{NONTP}, extending NTP's signal coverage along both dimensions through two auxiliary objectives. \textbf{TCL (Temporal Contrastive Learning)} uses a BYOL-style EMA teacher with InfoNCE to align hidden states against a $K$-step future trajectory in representation space. \textbf{TDL (Trans-Domain Learning)} mean-pools cross-domain hidden states and predicts through the shared prediction head, opening a second gradient pathway with no additional parameters. Both are discarded at inference: zero overhead.

On a four-domain Meituan industrial dataset (full ranking), NONTP achieves HR@10 +34.3\% over NTP and +18.3\% over MBGR. On the public Amazon Movie-Book-CDs benchmark, HR@10 +2.8\% and NDCG@10 +3.7\%. Online A/B tests confirm CTR +1.8\% and GMV +2.1\% (both $p < 0.01$). Ablation studies confirm each component contributes independently, with gradient conflict analyzed as a direction for future work.
\end{abstract}


\maketitle

\section{Introduction}

\subsection{Background}

Generative Recommendation (GR) has rapidly established itself as a leading paradigm for industrial-scale sequential recommendation. Inspired by the success of autoregressive language modeling, GR represents items as sequences of discrete semantic tokens and trains a Transformer to predict tokens via Next-Token Prediction (NTP). TIGER~\cite{tiger2023} pioneered this approach with RQ-VAE semantic IDs; HSTU~\cite{hstu2024} scaled it to industrial ranking with strong scaling-law properties; OneRec~\cite{onerec2025} further unified retrieval and ranking within a generative framework.

Yet a critical question remains: does NTP, designed for semantically continuous token sequences, transfer its full effectiveness to recommendation sequences? We argue it does not. Language sequences exhibit natural semantic continuity between adjacent tokens; recommendation sequences freely intermix items from fundamentally different semantic spaces.

Dissatisfaction with NTP has recently become a cross-domain theme in ML. In language generation, Nagarajan et al.~\cite{rollthedice2025} demonstrated NTP's myopia prevents it from completing tasks requiring implicit planning. At the representation level, Cola DLM~\cite{coladlm2026} and ELF~\cite{elf2026} abandon discrete tokens entirely, operating in continuous latent space. Our work joins this discussion at the training-signal level: we retain the discrete semantic ID framework and extend NTP's signal coverage through auxiliary objectives.

\subsection{Temporal Locality}

NTP trains a hidden state $\mathbf{h}_t$ to predict only the immediate next item $i_{t+1}$. The model is never given a direct supervised reason to encode information about $i_{t+2}, i_{t+3}, \dots, i_{t+K}$ into $\mathbf{h}_t$. While attention mechanisms theoretically allow long-range information flow, the loss function places zero pressure on the model to capture cross-step behavioral structure. We term this \textbf{temporal locality}: the training signal at each position is confined to a single future step, even though the user's true preference trajectory spans a broader horizon.

\textbf{Example.} Consider a user's weekly interaction sequence: ``food delivery $\to$ food delivery $\to$ food delivery $\to$ beauty salon,'' followed by ``beauty salon $\to$ group purchase $\to$ hotel booking.'' This trajectory clearly indicates a structural interest shift from instant consumption toward offline lifestyle services. Yet under NTP, at the second ``beauty salon'' position---the transition point---the hidden state is only trained to predict the immediate ``group purchase,'' completely missing the longer-range shift toward ``hotel booking'' two steps ahead. Only when supervision covers the $K=3$ window can the model perceive the systematic interest migration from ``beauty salon $\to$ group purchase $\to$ hotel booking.''

Temporal locality causes two practical problems: (1) \textbf{myopic bias}: the model memorizes high-frequency local transitions rather than capturing long-range interest evolution; (2) \textbf{noise sensitivity}: single next-item labels are extremely noisy (impulse purchases, gifts, misclicks), making the model prone to overfitting sporadic behaviors.

The use of RQ-VAE semantic IDs in GR further exacerbates temporal locality. In language modeling, adjacent tokens share a continuous semantic space, and gradient signals can implicitly carry semantic associations across positions. In GR, however, we argue that items encoded as discrete semantic ID token sequences can have entirely different codes even when semantically similar---providing no semantic bridge between consecutive tokens. This means NTP's myopic bias is even more severe in GR than in language: not only is the supervision window too short, but adjacent positions provide no implicit semantic continuity to compensate.

\subsection{Spatial Locality}

NTP is a \textbf{single-pathway} supervisor with respect to the shared prediction head $\mathbf{E}$: each target item's embedding $\mathbf{E}[i_{j+1}]$ receives gradient updates exclusively from the immediately preceding hidden state $\mathbf{h}_j$. While $\mathbf{h}_j$ aggregates information from all prior positions through self-attention, the \textit{gradient pathway} to $\mathbf{E}$ is structurally confined to a single input source. This constraint affects all targets uniformly, but its consequences are asymmetric across domains.

\textbf{Why sparse domains suffer disproportionately.} Consider a sparse-domain item $i_s$ in $\mathbf{E}$. Under NTP, $\mathbf{E}[i_s]$ receives one gradient update per occurrence as a next-item target---and sparse-domain items occur infrequently by definition. More critically, when $i_s$ does appear as a target, the single gradient pathway originates from $\mathbf{h}_j$, whose representation is shaped predominantly by dense-domain patterns (since dense-domain items dominate the training sequences). The sparse item embedding is thus trained on a diet of scarce, dense-domain-biased gradient signals.

TDL addresses this by opening a second gradient pathway to each target item embedding via cross-domain mean pooling, providing additional gradient signal beyond NTP's single-pathway constraint.

We term this structural limitation \textbf{spatial locality}: NTP confines each target's gradient pathway to a single source position, and provides no explicit mechanism for cross-domain context to participate in optimizing the shared prediction head.

MBGR~\cite{mbgr2026} partially addressed a related problem through Label Dynamic Routing (LDR), which replaces cross-domain labels with same-domain alternatives in label space---increasing the frequency with which sparse-domain items appear as prediction targets. LDR is an effective engineering solution, but it operates at the label level---choosing \textit{what} to predict---leaving the single-pathway constraint on $\mathbf{E}$ unchanged. This motivates our representation-space approach: rather than increasing target frequency, open additional gradient pathways to the shared head.

\subsection{NONTP Framework}

We propose \textbf{NONTP (Not Only NTP)}, extending NTP's training signal coverage through two auxiliary objectives trained jointly with the primary loss:

\textbf{TCL (Temporal Contrastive Learning).} TCL uses a BYOL-style EMA teacher with InfoNCE to contrast the current hidden state against a $K$-step future trajectory in representation space. Operating in hidden-state space rather than token-output space is deliberate: we argue that semantic IDs are discrete codes with no semantic continuity between adjacent item tokens, making representation-space alignment a more robust objective than token-sequence prediction for extending temporal supervision.

\textbf{TDL (Trans-Domain Learning).} For each position whose target belongs to domain $d$, TDL mean-pools hidden states from non-target-domain positions and predicts the target item through the \textbf{same shared prediction head} used by NTP, opening a second gradient pathway to each target item embedding. TDL introduces no additional head parameters, and mean pooling uniformly mixes cross-domain sources---smoothing domain-specific extremes while preserving shared patterns.

TCL extends supervision along the temporal axis, TDL along the domain axis, together providing dual-dimensional signal coverage.

Our contributions are:
\begin{enumerate}
\item We identify temporal locality as a fundamental limitation of NTP in recommendation, connecting it to the independently observed myopia of NTP in language modeling, and further characterize spatial locality in multi-domain sequences.
\item We propose TCL, a temporal contrastive objective with EMA teacher and InfoNCE that extends supervision to a $K$-step future trajectory in representation space.
\item We propose TDL, a cross-domain learning mechanism that mean-pools cross-domain context and predicts through the shared head, representing an evolution from label-space to representation-space approaches.
\item We validate NONTP on an industrial four-domain dataset and the public Amazon Movie-Book-CDs benchmark, with comprehensive ablation studies and online A/B results.
\end{enumerate}

\section{Preliminaries}

A generative recommendation model encodes a user interaction sequence $\mathcal{S}_u = [(i_1, d_1, t_1), \dots, (i_L, d_L, t_L)]$ through an HSTU backbone~\cite{hstu2024} to produce hidden states $\mathbf{H} = [\mathbf{h}_1, \dots, \mathbf{h}_L] \in \mathbb{R}^{L \times D}$, and predicts the next item through a shared prediction head $\mathbf{E} \in \mathbb{R}^{|\mathcal{I}| \times D}$:

\begin{equation}
P(i_{j+1} \mid i_1, \dots, i_j) = \text{softmax}(\mathbf{h}_j \cdot \mathbf{E}^T)
\end{equation}

The standard NTP objective maximizes the log-likelihood of observed next items:

\begin{equation}
\mathcal{L}_{\text{NTP}} = -\frac{1}{L-1} \sum_{j=1}^{L-1} \log P(i_{j+1} \mid i_1, \dots, i_j)
\end{equation}

In multi-domain recommendation, items partition into $M$ domains $\mathcal{D} = \{\mathcal{I}_1, \dots, \mathcal{I}_M\}$. User sequences freely intermix items from different domains, making cross-domain transitions a natural expression of user behavior rather than a data artifact.

\section{Problem Analysis}

\subsection{Temporal Locality}

NTP's myopia has been independently documented: Nagarajan et al.~\cite{rollthedice2025} demonstrated through algorithmic tasks that NTP can only learn features relevant to single-step prediction, and does not encode structure requiring longer-range planning. In recommendation, this same deficiency manifests concretely: the user's current preference state is encoded in a future behavioral trajectory, not any single next item, yet NTP's loss function never requires the model to capture this cross-step structure.

At each sequence position $j$, NTP asks $\mathbf{h}_j$ to do exactly one thing: predict $i_{j+1}$. While attention mechanisms theoretically allow $\mathbf{h}_j$ to attend to distant positions, the loss function places zero supervised pressure on $\mathbf{h}_j$ to encode patterns spanning $[j+2, j+K]$. This is not a claim that gradients \textit{cannot} flow---they can. It is a claim that the objective does not \textit{reward} learning long-range structure, and models naturally allocate capacity to what the loss directly measures.

\subsection{Spatial Locality: Single-Pathway Supervision}

NTP provides exactly one gradient pathway from $\mathbf{h}_j$ to the shared prediction head $\mathbf{E}$ for each target item $i_{j+1}$. This is a structural constraint, not a domain-specific artifact: regardless of whether $d_j = d_{j+1}$ or $d_j \neq d_{j+1}$, the target item embedding $\mathbf{E}[i_{j+1}]$ receives gradient updates from a single source position.

The practical consequence is asymmetric across domains. Sparse-domain items appear infrequently as targets, and even when they do, their single gradient pathway originates from $\mathbf{h}_j$---a hidden state whose representation is shaped predominantly by dense-domain training samples. A sparse item embedding in $\mathbf{E}$ thus receives scarce, biased gradient signals under NTP. Sparse domains are disproportionately affected because their within-domain transitions are few, yet their item embeddings in $\mathbf{E}$ have no alternative gradient source.

This motivates a representation-space remedy: provide a second, explicit gradient pathway to each target item embedding, sourced from cross-domain context rather than a single preceding position.

\subsection{Unified View of Prior Work}

Table~\ref{tab:unified} positions existing methods along the two locality axes.

\begin{table*}[t]
\centering
\caption{Unified view of prior work}
\label{tab:unified}
\begin{tabular}{lccp{4.5cm}}
\toprule
Method & Addresses & Level & Limitation \\
\midrule
NTP & -- & -- & Single-step supervision + single-pathway gradient \\
Future Data~\cite{futuredata2020} & Temporal (early) & Label & Session-level, not for GR \\
PinnerFormer~\cite{pinnerformer2022} & Temporal (industrial) & Label & Day-level skip labels, not position-level \\
SessionRec~\cite{sessionrec2025} & Temporal (partial) & Label & Session-dependent \\
LDR (MBGR)~\cite{mbgr2026} & Spatial (partial) & Label & Hard routing; requires domain labels \\
\midrule
\textbf{TCL (ours)} & \textbf{Temporal} & \textbf{Hidden state} & \textbf{EMA teacher adds training overhead; limited on short sequences} \\
\textbf{TDL (ours)} & \textbf{Spatial} & \textbf{Hidden state} & \textbf{Requires domain labels; limited without cross-domain context} \\
\bottomrule
\end{tabular}
\end{table*}

\begin{figure}[t]
\centering
\includegraphics[width=\columnwidth]{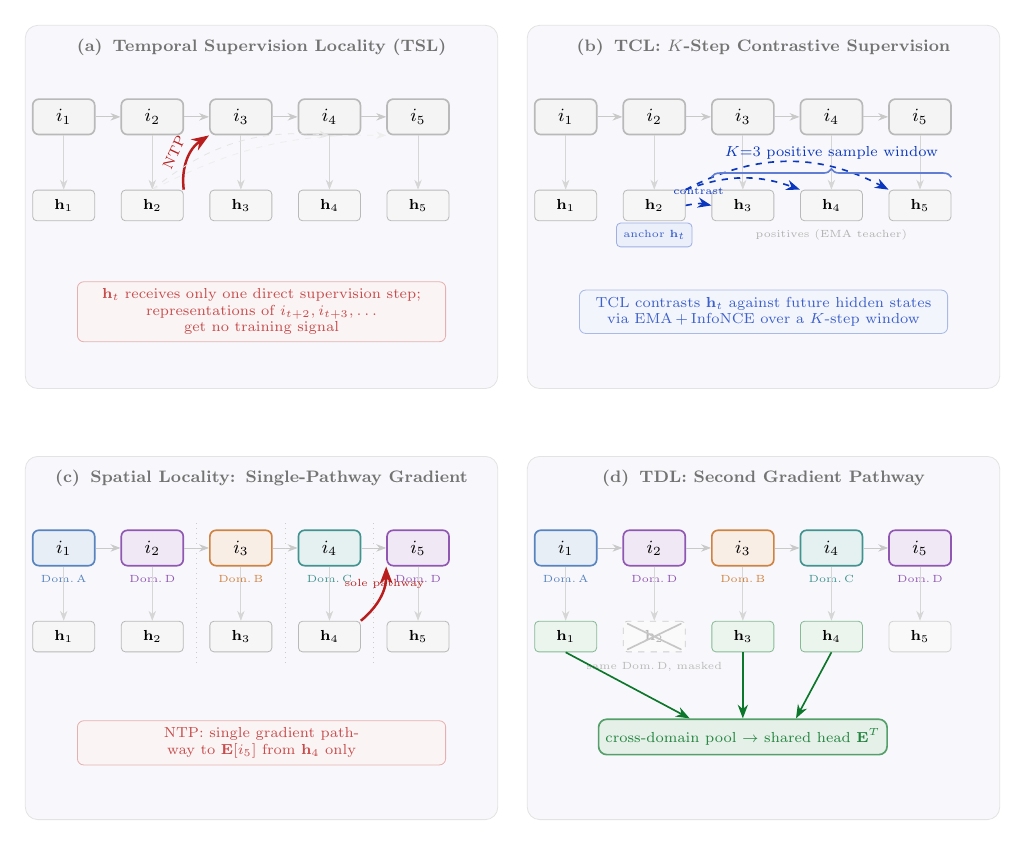}
\caption{Dual locality in NTP and NONTP's solutions. (a) Temporal locality: $\mathbf{h}_t$ receives direct supervision only from the immediate next item; future steps get no training signal. (b) TCL: EMA teacher + InfoNCE contrasts $\mathbf{h}_t$ against future hidden states over a $K$-step window. (c) Spatial locality: when predicting $i_5$ (Domain D), NTP provides only a single gradient pathway (from $\mathbf{h}_4$) to the target item embedding; information from positions $\mathbf{h}_1,\mathbf{h}_2,\mathbf{h}_3$---though present via attention---has no direct gradient path to $\mathbf{E}[i_5]$. (d) TDL: opens a second gradient pathway to $\mathbf{E}[i_5]$ by mean-pooling non-target-domain hidden states and predicting through the shared head.}
\label{fig:problem}
\end{figure}

Figure~\ref{fig:problem} illustrates the dual locality problem and NONTP's remediation.

\section{Method}

\subsection{Architecture Overview}
\label{sec:arch}

NONTP builds on an HSTU backbone. Training jointly optimizes three losses via simple weighted summation:

\begin{equation}
\mathcal{L} = \mathcal{L}_{\text{NTP}} + \lambda_{\text{tcl}} \mathcal{L}_{\text{TCL}} + \lambda_{\text{tdl}} \mathcal{L}_{\text{TDL}}, \quad \lambda_{\text{tcl}} = \lambda_{\text{tdl}} = 0.1
\end{equation}

\begin{figure}[t]
\centering
\includegraphics[width=\columnwidth]{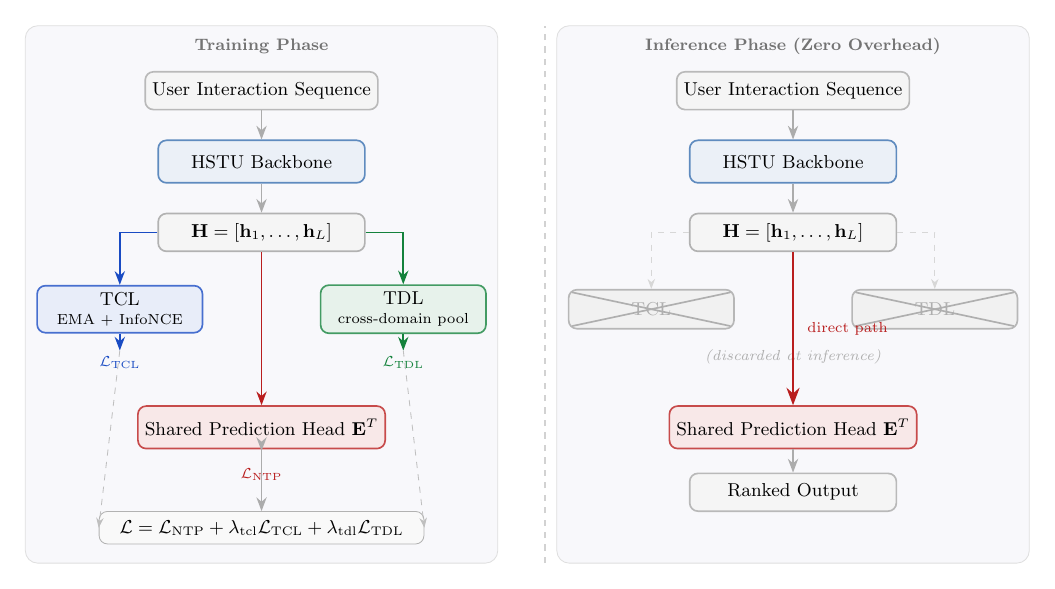}
\caption{Training (left) and Inference (right). During training, three losses are computed from the HSTU hidden states and summed directly. At inference, all auxiliary components (EMA teacher, TCL predictors, TDL pooling) are discarded; NTP logits are computed through a direct path from $\mathbf{H}$ to the shared prediction head. The deployed model is identical in structure, parameters, and inference cost to the NTP baseline---zero inference overhead.}
\label{fig:arch}
\end{figure}

At inference, all auxiliary components (EMA teacher, TCL predictors, TDL pooling) are discarded. The deployed model is a standard HSTU backbone with the shared prediction head---identical in structure, parameters, and inference cost to the NTP baseline. Figure~\ref{fig:arch} illustrates the training and inference architectures.

\subsection{TCL: Temporal Contrastive Learning}

TCL's core insight: \textbf{don't predict what the future item is; let $\mathbf{h}_t$ perceive the distribution of future states.} Following CPC~\cite{cpc2018} and BYOL~\cite{byol2020}, TCL uses an EMA teacher to produce stable target representations and InfoNCE to align predictions against them.

\textbf{EMA Teacher.} An exponential moving average copy of the backbone serves as teacher:
\begin{equation}
\theta_{\text{ema}} \leftarrow m \cdot \theta_{\text{ema}} + (1 - m) \cdot \theta_{\text{online}}, \quad m = 0.999
\end{equation}
producing stable target states $\mathbf{h}^{\text{ema}}_1, \dots, \mathbf{h}^{\text{ema}}_L$.

\textbf{Future-step predictors.} For each offset $k \in \{1, \dots, K\}$, a linear predictor $g_k: \mathbb{R}^D \to \mathbb{R}^D$ maps $\mathbf{h}_j$ to a forward prediction $\mathbf{z}^{(k)}_j = g_k(\mathbf{h}_j)$.

\textbf{InfoNCE loss.} For each $k$ and each valid position $j$ ($j+k \leq L$):
\begin{equation}
\begin{split}
\mathcal{L}_{\text{TCL}} = -\frac{1}{K} \sum_{k=1}^{K} \frac{1}{|\mathcal{P}_k|} \sum_{j \in \mathcal{P}_k} \log \\
\frac{\exp(\text{sim}(\mathbf{z}^{(k)}_j,\, \mathbf{h}^{\text{ema}}_{j+k}) / \tau)}{\displaystyle\sum_{(s',j') \in \mathcal{B}_k} \exp(\text{sim}(\mathbf{z}^{(k)}_j,\, \mathbf{h}^{\text{ema},(s')}_{j'+k}) / \tau)}
\end{split}
\end{equation}
where $\mathcal{P}_k = \{j : j+k \leq L\}$; $\mathcal{B}_k$ is the set of all valid (sequence~$s'$, position~$j'$) pairs in the batch with $j'{+}k \leq L_{s'}$; $\mathbf{h}^{\text{ema},(s')}_{j'+k}$ denotes the EMA hidden state at position $j'{+}k$ in sequence $s'$; and $\text{sim}$ is cosine similarity with temperature $\tau$. Positive pairs are constructed independently for each offset $k$. Negatives are drawn exclusively from \emph{other sequences} in the batch; same-sequence positions are excluded to avoid false negatives arising from within-sequence temporal correlations.

\textbf{Inference.} EMA teacher and all predictors are discarded. Zero overhead.

\subsection{TDL: Trans-Domain Learning}

TDL addresses spatial locality by opening a second gradient pathway to each target item embedding in $\mathbf{E}$. Each item $i$ belongs to a domain $d(i)$ determined by dataset metadata (e.g., Amazon's Movie/Book/CDs categories; Meituan's business-line tags); we write $d_{j+1} \equiv d(i_{j+1})$ for the target item's domain. For a position $j$, define the cross-domain context set:
\begin{equation}
\mathcal{C}_j = \{l < j : d_l \neq d_{j+1}\}
\end{equation}

When $\mathcal{C}_j$ is non-empty, compute a cross-domain representation by mean pooling:
\begin{equation}
\mathbf{z}^{\text{tdl}}_j = \frac{1}{|\mathcal{C}_j|} \sum_{l \in \mathcal{C}_j} \mathbf{h}_l
\end{equation}

This representation predicts the target item through the \textbf{same shared prediction head} used by NTP:
\begin{equation}
\hat{P}(i_{j+1} \mid \mathcal{C}_j) = \text{softmax}(\mathbf{z}^{\text{tdl}}_j \cdot \mathbf{E}^T)
\end{equation}

The TDL loss is:
\begin{equation}
\mathcal{L}_{\text{TDL}} = -\frac{1}{|\mathcal{J}_{\text{tdl}}|} \sum_{j \in \mathcal{J}_{\text{tdl}}} \log \hat{P}(i_{j+1} \mid \mathcal{C}_j)
\end{equation}
where $\mathcal{J}_{\text{tdl}} = \{j : |\mathcal{C}_j| > 0\}$. Since $d_{j+1}$ is static metadata, $\mathbf{z}^{\text{tdl}}_j$ is computed exclusively from hidden states at positions $l < j$; no content information about $i_{j+1}$ is introduced.

TDL opens a second gradient pathway to each target item embedding: NTP provides one gradient update per target occurrence (from $\mathbf{h}_j$), while TDL provides an additional update from the cross-domain pooled representation $\mathbf{z}^{\text{tdl}}_j$. For sparse-domain items that appear infrequently as targets, this doubles the gradient updates per occurrence---and the second pathway's input is sourced predominantly from dense-domain positions. Mean pooling uniformly mixes contributions from different domains: single-domain extremes are smoothed, while cross-domain shared signals are retained. TDL can be seen as a representation-space evolution of LDR: label-space hard routing $\to$ representation-space soft aggregation, sharing the prediction head with NTP.

\subsection{TCL and TDL: Shared Design Principles}

TCL and TDL share a common design principle: both operate in hidden-state representation space, both avoid introducing custom label definitions, both are orthogonal to the primary NTP objective, and both are discarded at inference. TCL extends temporal coverage; TDL extends cross-domain coverage (total loss in §\ref{sec:arch}, default weights $\lambda_{\text{tcl}} = \lambda_{\text{tdl}} = 0.1$).

\section{Related Work}

\begin{sloppypar}
\textbf{Generative Recommendation.} TIGER~\cite{tiger2023}, HSTU~\cite{hstu2024}, and OneRec~\cite{onerec2025} established the GR paradigm. MBGR~\cite{mbgr2026} extended it to multi-business settings with BID, MBP, and LDR; NONTP reuses BID\slash MBP infrastructure and positions LDR as a label-space predecessor to TDL. SessionRec~\cite{sessionrec2025} extended prediction to session-level targets.

\textbf{Limitations of NTP.} Nagarajan et al.~\cite{rollthedice2025} demonstrated NTP myopia on algorithmic tasks. APAO~\cite{apao2026} identified train-inference mismatch in GR beam search. Zhang et al.~\cite{token2item2026} proposed item-aware attention. Ding et al.~\cite{grgeneralize2026} analyzed GR memorization vs. generalization. In LLMs, Mahajan et al.~\cite{beyondmtp2026} used future summaries; Zhong et al.~\cite{consistentwm2026} analyzed MTP gradient bias. At the representation level, Cola DLM~\cite{coladlm2026} and ELF~\cite{elf2026} advocate continuous latent spaces over discrete tokens---a fundamentally different critique from ours. NONTP operates at the training-signal level: it retains the discrete semantic ID framework and extends NTP's signal coverage through auxiliary objectives, complementing rather than competing with representation-level approaches.
\end{sloppypar}

\textbf{Future Supervision in Recommendation.} Future Data Helps Training~\cite{futuredata2020} introduced future context for session-based rec. PinnerFormer~\cite{pinnerformer2022} used 14-day future behaviors as skip labels. Oracle4Rec~\cite{oracle4rec2025} proposed oracle-guided preference modeling. TCL differs in operating on GR hidden states with contrastive learning rather than discriminative embeddings with direct supervision.

\textbf{Contrastive and Self-Supervised Representation Learning.} BYOL~\cite{byol2020} demonstrated that stable self-supervised targets can be learned through an EMA teacher without explicit negative sampling, while CPC~\cite{cpc2018} introduced InfoNCE for predictive representation learning over future steps. TCL can be understood as adapting the BYOL + CPC design pattern to generative recommendation: EMA teacher for stable future-state targets, and InfoNCE for multi-step temporal alignment in hidden-state space.

\begin{sloppypar}
\textbf{Distinction from CDR.} Cross-Domain Recommendation (CDR) transfers knowledge across separate datasets with explicit source\slash target domains. NONTP operates on a single dataset where user sequences naturally intermix multiple domains---there is no cross-dataset transfer objective.
\end{sloppypar}

\section{Experiments}

\subsection{Datasets}

\textbf{Meituan Industrial Dataset.} Covers one year of user interaction sequences across four business domains (A/B/C/D) on the Meituan platform. The training set covers 38.3M users and 54.9M items. Domain density varies substantially: Domain A (dominant), B (medium-density), C and D (sparse). Average sequence length is 280. Evaluation: \textbf{full ranking} over the complete item corpus, reporting HR@10 per domain and overall (\textbf{All}: sample-weighted average across domains).

\textbf{Amazon Movie-Book-CDs.} Constructed from the Amazon review corpus~\cite{amazon2015}: Movie, Book, and CDs domains. User sequences are built by merging per-user interaction histories across categories, ordered strictly by timestamp, so cross-domain transitions reflect the user's actual temporal behavior. Test set: 1,241,628 samples (Movie: 480K; Book: 720K; CDs: 42K), 3.2M total items. Average sequence length is 12. Evaluation: \textbf{fixed negative sampling} (999 candidates, seed=42). Report HR@1/5/10 and NDCG@5/10.

\subsection{Baselines}

SASRec~\cite{sasrec2018}, TIGER~\cite{tiger2023}, NTP (HSTU), and MBGR~\cite{mbgr2026}. NONTP ablation variants: +TCL (NTP + TCL), +TDL (NTP + TDL), and NONTP (full).

\subsection{Implementation}

Adam optimizer, learning rate grid-searched from $\{1\text{e-}4, 3\text{e-}4, 5\text{e-}4\}$, max sequence length 1500, early stopping on validation NDCG@10 (patience=5), averaged over 3 seeds (std $\sigma \approx 0.0006$ for industrial All HR@10). TCL: $K{=}3$, EMA $m{=}0.999$, $\tau{=}0.07$. $\lambda_{\text{tcl}} = \lambda_{\text{tdl}} = 0.1$.

Training cost: NONTP adds an EMA teacher (full backbone copy), $K$ linear predictor heads, InfoNCE pairwise computation, and TDL cross-domain pooling. Training time increases $\sim$20--25\%, GPU memory $\sim$40\% vs.\ NTP. Inference: zero overhead.

\subsection{Industrial Results}

Table~\ref{tab:main} reports full-ranking HR@10. NONTP consistently outperforms all baselines across all domains (All HR@10 0.0485, +34.3\% vs.\ NTP, +18.3\% vs.\ MBGR). Sparse domains (C: +0.0114, D: +0.0119 absolute) show larger absolute gains than the dense domain (A: +0.0058), consistent with the dual-pathway mechanism: sparse-domain item embeddings in $\mathbf{E}$ receive the fewest NTP gradient updates per unit of training, and thus benefit most from TDL's second gradient pathway---which is predominantly sourced from dense-domain positions. We note that Domain A's lower NTP baseline (0.0222) may reflect its larger item catalog making full ranking inherently harder; cross-domain absolute gain comparison should account for catalog size differences.

\begin{table}[b]
\centering
\footnotesize
\setlength{\tabcolsep}{3pt}
\caption{Meituan industrial results (HR@10, full ranking). Best in bold.}
\label{tab:main}
\begin{tabular}{lrrrrr}
\toprule
Method & All & A & B & C & D \\
\midrule
SASRec & 0.0192 & 0.0218 & 0.0101 & 0.0178 & 0.0269 \\
TIGER & 0.0202 & 0.0221 & 0.0128 & 0.0180 & 0.0278 \\
NTP & 0.0361 & 0.0222 & 0.0488 & 0.0351 & 0.0371 \\
MBGR & 0.0410 & 0.0252 & 0.0554 & 0.0398 & 0.0421 \\
\midrule
+TCL & 0.0445 & 0.0268 & 0.0595 & 0.0438 & 0.0460 \\
+TDL & 0.0458 & 0.0273 & 0.0612 & 0.0448 & 0.0472 \\
\textbf{NONTP} & \textbf{0.0485} & \textbf{0.0280} & \textbf{0.0630} & \textbf{0.0465} & \textbf{0.0490} \\
\bottomrule
\end{tabular}
\end{table}

The industrial-public gain gap (+34.3\% vs.\ +2.8\%) reflects multiple structural differences: full-ranking vs.\ 999-candidate negative sampling, four vs.\ three domains, and---most prominently---sequence length (280 vs.\ 12 steps). Longer sequences amplify both temporal locality (myopia accumulates over more steps) and cross-domain transitions, making the industrial dataset a stress test for dual locality while the public dataset validates generalization. The consistent rank ordering NONTP $>$ TDL $>$ TCL $>$ NTP across both datasets confirms that gains stem from the method rather than dataset-specific factors.

\subsection{Public Dataset Results}

Table~\ref{tab:amazon} reports Amazon results. NONTP achieves HR@10 +2.8\% and NDCG@10 +3.7\% overall. Table~\ref{tab:amazon-domain} reports per-domain results. Gains are consistent across all three domains (Movie: +0.0164, Book: +0.0060, CDs: +0.0027 absolute HR@10). The CDs domain (42K samples, 3.4\% of test set) shows the smallest improvement, which differs from the industrial pattern where sparse domains benefit most. We attribute this to the extremely short Amazon sequences (avg.\ 12 steps): with few total transitions, the single-pathway constraint has limited impact, and the benefit of a second gradient pathway is correspondingly small.

\begin{table}[!h]
\centering\scriptsize
\caption{Amazon Movie-Book-CDs results (999 candidates, seed=42).}
\label{tab:amazon}
\begin{tabular}{lrrrr}
\toprule
Method & HR@1 & HR@5 & HR@10 & $\Delta$HR@10 \\
\midrule
NTP    & 0.1458 & 0.2796 & 0.3455 & -- \\
+TCL   & 0.1480 & 0.2836 & 0.3498 & +1.2\% \\
+TDL   & 0.1505 & 0.2864 & 0.3532 & +2.2\% \\
\textbf{NONTP} & \textbf{0.1534} & \textbf{0.2895} & \textbf{0.3553} & \textbf{+2.8\%} \\
\bottomrule
\end{tabular}
\par\vspace{2pt}
\begin{tabular}{lrrr}
\toprule
Method & NDCG@5 & NDCG@10 & $\Delta$NDCG@10 \\
\midrule
NTP    & 0.2158 & 0.2371 & -- \\
+TCL   & 0.2189 & 0.2405 & +1.4\% \\
+TDL   & 0.2217 & 0.2432 & +2.6\% \\
\textbf{NONTP} & \textbf{0.2246} & \textbf{0.2459} & \textbf{+3.7\%} \\
\bottomrule
\end{tabular}
\end{table}

\begin{table}[!h]
\centering\scriptsize
\caption{Amazon per-domain results (HR@10 / NDCG@10).}
\label{tab:amazon-domain}
\begin{tabular}{lrrr}
\toprule
Method & Movie & Book & CDs \\
\midrule
\multicolumn{4}{l}{\textit{HR@10}} \\
NTP & 0.4118 & 0.3058 & 0.2657 \\
\textbf{NONTP} & \textbf{0.4282} & \textbf{0.3118} & \textbf{0.2684} \\
\midrule
\multicolumn{4}{l}{\textit{NDCG@10}} \\
NTP & 0.2809 & 0.2123 & 0.1598 \\
\textbf{NONTP} & \textbf{0.2947} & \textbf{0.2182} & \textbf{0.1625} \\
\bottomrule
\end{tabular}
\end{table}

\subsection{Ablation Studies}

Table~\ref{tab:ablation} reports absolute All HR@10 for all ablation variants on the industrial dataset (base: NONTP = 0.0485). For TDL, the \textit{no-pool} variant makes one independent prediction per cross-domain position and averages the losses (no aggregation); the remaining rows compare pooling strategies.
\begin{table}[t]
\centering
\caption{Ablation results (industrial All HR@10; base NONTP = 0.0485).}
\label{tab:ablation}
\scriptsize
\begin{tabular}{lcc}
\toprule
Variant & HR@10 & $\Delta$ \\
\midrule
\multicolumn{3}{l}{\textit{TCL ablation (TDL fixed at full)}} \\
NONTP (full) & 0.0485 & — \\
$-$ EMA teacher & 0.0470 & $-$3.1\% \\
$K{=}1$ & 0.0475 & $-$2.1\% \\
$K{=}5$ & 0.0480 & $-$1.0\% \\
$-$ predictor MLPs & 0.0477 & $-$1.6\% \\
\midrule
\multicolumn{3}{l}{\textit{TDL pooling strategy (TCL fixed at full)}} \\
NONTP (full) & 0.0485 & — \\
no-pool (per-pos.\ $\mathbf{h}$) & 0.0478 & $-$1.4\% \\
max-pool & 0.0483 & $-$0.4\% \\
attention-pool & 0.0485 & $\approx$0\% \\
\bottomrule
\end{tabular}
\end{table}

\begin{sloppypar}
\textbf{Hyperparameter sensitivity.} Temperature $\tau$ is stable in $[0.05, 0.10]$; too small ($\tau{=}0.01$) causes instability, too large ($\tau{=}0.20$) weakens the contrastive signal. The loss weights $\lambda_{\text{tcl}}, \lambda_{\text{tdl}}$ are stable in $[0.05, 0.20]$ with main results varying under 1\%. $m{=}0.999$ is optimal; $m{=}0.99$ performs 1.8\% worse.
\end{sloppypar}

\textbf{Gradient conflict analysis.} Directly summing three losses introduces the risk of gradient interference: the objectives can pull shared backbone parameters in conflicting directions. All reported results use simple weighted summation at $\lambda_{\text{tcl}} = \lambda_{\text{tdl}} = 0.1$; we verified empirically that this setting places the system in a cooperative regime, and that increasing either weight beyond $\approx$0.3 degrades HR@10, consistent with gradient conflict dominating at larger scales. We also conducted a preliminary experiment applying PCGrad-style gradient surgery~\cite{pcgrad2020}---projecting TCL/TDL gradient components that conflict with the NTP gradient onto orthogonal directions before summation---and observed a modest additional improvement. However, projection only addresses directional conflict, not magnitude mismatch (auxiliary losses can dominate early training). Improved multi-objective gradient coordination remains an open direction.

\subsection{Online A/B Test}

We deployed NONTP on the Meituan DSP platform's recall stage, running an A/B test (5\% traffic each, 14 days). NONTP achieved CTR +1.8\% and GMV +2.1\% (both $p < 0.01$), confirming offline-to-online consistency.

\section{Discussion}

\subsection{Relationship to MBGR}

NONTP builds on MBGR's BID and MBP components for multi-domain item representation and alignment. MBGR's LDR addresses cross-domain sparsity at the label level through hard routing; TDL extends this to representation-space soft aggregation, sharing the prediction head with NTP. TCL addresses the orthogonal temporal dimension absent from MBGR. The two works represent a progression from label-space engineering to representation-space training signal design.

\subsection{Limitations}

NONTP has the following limitations: (1) TDL requires domain labels; in settings without explicit annotations, domains must be inferred via clustering. (2) TCL's EMA momentum scheduling is sensitive; adaptive scheduling is an open question. (3) TDL requires at least two cross-domain positions, limiting effectiveness on very short sequences ($<$10 steps). Future work includes adaptive domain label inference, dynamic $K$ selection, improved gradient conflict resolution, and extending the signal-coverage framework to broader sequence modeling scenarios.

\section{Conclusion}

We identified temporal locality (single-step supervision) and spatial locality (single-pathway gradient to the shared prediction head) as two structural training signal limitations of NTP in generative recommendation. NONTP addresses them through TCL (EMA teacher + InfoNCE over $K$-step future trajectory) and TDL (cross-domain mean pooling + shared prediction head), both trained jointly with the NTP loss and discarded at inference. Experiments on an industrial four-domain dataset and Amazon Movie-Book-CDs demonstrate consistent, substantial improvements. The dual-locality framework offers a principled lens for diagnosing and addressing training signal insufficiency in autoregressive recommendation models.

\bibliographystyle{ACM-Reference-Format}
\bibliography{references}

\end{document}